# Thermonuclear Fusion with the Sheared Flow Stabilized Z-Pinch

F. Winterberg and L. F. Wanex

*University of Nevada Reno, Department of Physics / MS 372, Reno, Nevada, 89557, USA*

**ABSTRACT**

Two basic approaches to producing thermonuclear fusion with a sheared flow stabilized z-pinch are considered. One consists of heating the entire length of the z-pinch column to the required temperatures. The other basic approach considered here involves the concept of fast ignition. A localized "hot-spot" is produced under the proper conditions to ignite a thermonuclear burn wave in the z-pinch plasma. Here we demonstrate that sheared flow stabilization is more efficient in the fast-ignition method with isentropic compression then in a z-pinch where the entire plasma column is heated.

**Keywords** z-pinch, sheared flow, fast ignition

## I. INTRODUCTION

Some of the earliest attempts to realize controlled nuclear fusion were based on the z-pinch, where a large current is discharged through a column of deuterium-tritium (DT) gas, to compress and heat the DT to the ignition temperature of $T \geq 10$ keV. Subsequently, numerous z-pinch configurations have been studied to reach the conditions for thermonuclear ignition and burn. Here we consider two basic approaches to this problem. In the first, the entire length of the z-pinch column is heated to the required temperature for ignition. Conceivably, this can be done by resistive heating and compression of the stagnated z-pinch plasma [1, 2], or by isentropically compressing the z-pinch with a programmed fast rising current [3]. In a variation of this method a large current is discharged over a frozen DT fiber [4]. Laser heating of the entire pinch column has also been proposed to raise the temperature to ignition [5-7]. All these approaches suffer from the instability of the pinch discharge. To overcome this problem the stabilization by a sheared axial flow along the pinch discharge channel has been studied [8]. This stabilization concept was first introduced by one of us, proposing to stabilize the pinch discharge by shooting through its core a high-velocity needle-like projectile [9].

The second approach involves the concept of fast ignition. There, a "hot-spot" is created in the pinch discharge channel, to ignite a thermonuclear burn wave propagating down the entire length of the channel. Various methods have been proposed to create a hot-spot. One, is to make it in a localized region by the sausage ($m = 0$) instability [10, 11]. Another method is to focus a high energy electron- or a laser-beam on one end of the pinch column. This method has been proposed as early as 1969 [12-14]. We believe this second method is to be preferred over the $m = 0$ instability hot-spot generation, because as we will show, it permits to eliminate this instability by sheared axial and rotational flow. If the pinch current is large enough, it can confine the charged fusion products within the channel, a necessary condition for detonation wave propagation down the pinch discharge channel [15, 16].

The z-pinch is very vulnerable to instabilities, limiting its lifetime. Means of mitigating and/or suppressing macro- and micro-instabilities, turbulence, and transport in plasmas are essential elements of magnetic confinement research. For a z-pinch plasma, the fastest growing instabilities are the sausage $m = 0$ and kink $m = 1$ modes. If the pinch discharge channel is imploded, the Rayleigh-Taylor instability must also be suppressed .

Experimental [17], computational [18] and analytical [19] studies have demonstrated that sheared poloidal or azimuthal flows [20] are particularly effective control knobs in tokamak and stellarator plasmas. In addition to the stabilizing effect of an axial magnetic field [1], sheared axial and/or azimuthal flows have also been shown to be effective in controlling the linear [21-24] and nonlinear [25, 26] developments of sausage and kink instabilities in Z-pinch plasmas. Here we focus on the effect of sheared helical and azimuthal flows and demonstrate that this method of suppressing MHD instabilities is particularly well suited for fast-ignition.

The remainder of this paper is organized as follows: In section II we present analytical results that azimuthal flow can reduce the growth of z-pinch instabilities. Section III contains numerical results supporting the concept that the sheared azimuthal flow in a vortex can reduce the growth of the kink instability. In section IV we do the same for sheared helical flow. Section V, summarizes the results.

## II. AZIMUTHAL FLOW

For a uniformly rotating plasma with azimuthal velocity $v > v_A = B/\sqrt{4\pi\rho}$, the plasma is stable with regard to magnetohydrodynamic disturbances, as can be seen from the magnetohydrodynamic equation of motion ($p$ plasma pressure, $\mathbf{j}$ electric current density)

$$\frac{\partial \mathbf{v}}{\partial t} = -\frac{1}{\rho}\nabla p - \frac{1}{2}\nabla v^2 + \mathbf{v}\times(\nabla\times\mathbf{v}) + \frac{1}{\rho c}\mathbf{j}\times\mathbf{B}. \tag{1}$$

For a uniformly rotating plasma one has

$$\mathbf{v} = \boldsymbol{\omega}\times\mathbf{r} \tag{2}$$

where $\boldsymbol{\omega}$ is directed along the axis of rotation. Inserting (2) into (1) one has

$$\frac{\partial \mathbf{v}}{\partial t} = -\frac{1}{\rho}\nabla p - \omega^2 \mathbf{r} - 2\boldsymbol{\omega}\times\mathbf{v} + \frac{1}{\rho c}\mathbf{j}\times\mathbf{B}. \tag{3}$$

The magnetohydrodynamic instabilities come from the last term on the right had side of (3). For $v > v_A$, where $(1/2)\rho v^2 > B^2/8\pi$, the magnetic field lines are forced to align themselves with the lines of the super-Alfvén velocity plasma flow. Then, if likewise $\mathbf{j}$ aligns itself with $\boldsymbol{\omega}$, one has $\mathbf{j}\perp\mathbf{B}$, since $\boldsymbol{\omega}\perp\mathbf{v}$, and $\boldsymbol{\omega}\perp\mathbf{B}$. With $\boldsymbol{\omega} = (1/2)\nabla\times\mathbf{v}$ and $\mathbf{j} = (c/4\pi)\nabla\times\mathbf{B}$, the magnetic forces are overwhelmed by the fluid flow forces if

$$|\mathbf{v} \times (\nabla \times \mathbf{v})| \gg \left|\frac{\mathbf{B} \times (\nabla \times \mathbf{B})}{4\pi\rho}\right| \qquad (4)$$

or if **v** = const **B**, simply if

$$v > v_A. \qquad (5)$$

A uniformly rotating fluid can also be viewed as a lattice of potential vortices. Here too, it is easy to understand why a magnetized plasma is stable for v > v$_A$, where v is the fluid velocity of the potential vortex. To prove stability, we compare a linear pinch discharge where outside the pinch column $\nabla \times \mathbf{B} = \mathbf{0}$, with a potential vortex where outside the vortex core $\nabla \times \mathbf{v} = \mathbf{0}$ (see Fig. 1). Because of $\nabla \times \mathbf{B} = \mathbf{0}$, the magnetic field strength gets larger with decreasing distance from the center of curvature of the magnetic field lines. For $\nabla \times \mathbf{v} = \mathbf{0}$, the same is true for the velocity of a potential vortex. But whereas in the pinch discharge a larger magnetic field means a larger magnetic pressure, a larger fluid velocity means a smaller pressure by virtue of Bernoulli's theorem. Therefore, whereas a pinch column is unstable with regard to its bending, the opposite is true for a line vortex. Thus, a pinch column can be stabilized by placing it into a vortex provided $v > v_A$. What is true for the *m* = 1 kink instability is also true for the *m* = 0 sausage instability by the conservation of circulation

$$Z = \oint \mathbf{v} \, d\mathbf{r} = const. \qquad (6)$$

which implies that upon its pinching inside a vortex, $v \propto 1/r$ with $\rho v^2 \propto 1/r^2$, rising in the same proportion as the magnetic pressure, stabilizing the pinch against the *m* = 0 instability for $v > v_A$. And because of the centrifugal force, the vortex also stabilizes the plasma against the Rayleigh-Taylor instability.

With the magnetic field in the azimuthal and the current in the axial direction, a uniformly, super-Alfvén velocity, rotating plasma can be viewed as a large, vortex stabilized z-pinch.

### III. SHEARED AZIMUTHAL FLOW

Adding shear to the azimuthal velocity is also effective at reducing the growth of instabilities. Velocity shear stabilization is primarily a phase mixing process that disrupts the growth of unstable modes [19]. A linear stability analysis of the effect of sheared helical flow on the kink mode provides support for this concept. The MHD equations used are

$$\frac{\partial p}{\partial t} = -(\mathbf{v} \cdot \nabla) p - \gamma\, p \nabla \cdot \mathbf{v} \tag{7}$$

$$\frac{\partial \rho}{\partial t} = -\nabla \cdot (\rho \mathbf{v}) \tag{8}$$

$$\frac{\partial \mathbf{B}}{\partial t} = \nabla \times (\mathbf{v} \times \mathbf{B}) \tag{9}$$

$$\frac{\partial \mathbf{v}}{\partial t} = -(\mathbf{v} \cdot \nabla)\mathbf{v} - \frac{\nabla p}{\rho} + \frac{1}{4\pi\rho}(\nabla \times \mathbf{B}) \times \mathbf{B} \tag{10}$$

where $p$ is the pressure, $\mathbf{v}$ the velocity, $\rho$ the density, $\mathbf{B}$ the magnetic field, and $\gamma$ the adiabatic constant. The linearized equations are solved numerically with a generic two-step predictor-corrector, second-order accurate space and time-centered advancement scheme [23]. The problem is treated by introducing perturbations into the plasma equilibrium state and following their linear development in time. All perturbations have the mathematical form $\xi(r)e^{i(m\phi + k_z z - \omega t)}$ where $m$ is the azimuthal mode number, $k_z$ is the axial wave number and $\omega$ the frequency. Initially the growth rates of the perturbed plasma variables are uncorrelated, however, after several growth times the solution converges to the fastest growing unstable mode.

The plasma equilibrium profile is obtained from the radial force balance equation

$$\frac{\partial}{\partial r}\left(p_0 + \frac{B_{0\theta}^2}{8\pi} + \frac{B_{0z}^2}{8\pi}\right) + \frac{B_{0\theta}^2}{4\pi r} = \frac{mn_0 v_{0\theta}^2}{r}. \qquad (11)$$

By assuming that the electron and ion temperatures are equal we obtain $n_0 = p_0/mv_{th}^2$ where $v_{th}$ is the thermal velocity. In considering the effects of sheared flow alone the axial magnetic field is set to zero $B_{0z} = 0$. Once the magnetic field and velocity are entered into (11) the equilibrium pressure profile can be obtained numerically.

To test the stabilizing nature of sheared azimuthal flow we have performed a stability analysis on several different forms of sheared azimuthal flow. Figure 4 shows the kink instability growth rates versus axial wave number $k_z r_0$ for each of several different flow profiles. The profile with sheared azimuthal flow in the form $v_\phi = 0.1 v_A / \sqrt{0.01 + r^2}$ reduces growth rates of the kink instability more efficiently than the others ($v_A$ is the Alfvén velocity). This azimuthal flow has the form

$$\frac{0.1 v_A}{\sqrt{0.01 + r^2}} = \frac{0.1 v_A}{r\sqrt{1 + \frac{0.01^2}{r^2}}} \sim \frac{0.1 v_A}{r}, \qquad (12)$$

which for $0.01^2/r^2 < 1$ is a potential vortex.

A possible method for introducing sheared azimuthal flow into a z-pinch implosion is through the use of a magnetic cusp produced by opposing solenoids. By imploding a wire array in the center of the cusp, the axisymmetric radial magnetic field will cause the imploding plasma to rotate by the $\boldsymbol{j} \times \boldsymbol{B} = [0, j_z B_r, 0]$ force. An efficient load (see figure 3 below) for this configuration can be created by placing the opposing solenoids in series with the cylindrical wire array.

**IV. SHEARED HELICAL FLOW STABILIZATION OF THE Z-PINCH**

Axial sheared flow has been shown to reduce the linear growth rates of the $m = 0$ and $m = 1$ instabilities in z-pinch plasmas for some equilibrium profiles [9, 21-23]. The principal result established in these works is that sheared axial flow stabilizes the sausage and kink instabilities when the shear exceeds a threshold. It has also been established that sheared rotational flow reduces MHD instabilities as well [27, 28], and that the $\mathbf{v}_E = \mathbf{E} \times \mathbf{B} / B^2$ rotational flow with shear has also a stabilizing influence on MHD instabilities in Tokomak plasmas.

Since both sheared axial and azimuthal flow can reduce the growth of z-pinch instabilities, this evokes the possibility of combining azimuthal and axial flow to produce sheared helical flow as a stabilizing mechanism for z-pinch implosions. It has been suggested that sheared helical flow is a more efficient means of mitigating z-pinch instabilities than the application of axial or azimuthal sheared flow alone [29, 30]. This possibility is of great interest because of the profound implication a stable, high-density z-pinch would have for magnetic confinement thermonuclear fusion [21].

A vortex combined with sheared axial flow can reduce the growth of z-pinch instabilities. This can be seen as follows: The density of the destabilizing force acting on the pinch plasma is

$$f_H = \frac{B^2}{4\pi\lambda} \tag{13}$$

where $B$ is the magnetic field and $\lambda$ is the wavelength of a disturbance by which the pinch plasma is displaced in the radial direction. But by pushing against the wall of the vortex the pinch plasma is then also subject to the buoyancy force

$$f_B = \left(1 - \frac{\rho_p}{\rho}\right)\frac{dp}{dr} \tag{14}$$

where $\rho_p$ is the density of the pinch plasma and $\rho$ is the density of the imploding vortex. Stability requires that

$$f_H + f_B < 0. \tag{15}$$

Using the assumption that the solutions to equations (7-10) have the form

$$u r = F(t) \tag{16}$$

$$w r = G(t) \tag{17}$$

it can be shown that stability can be assured for all wavelengths down to $\lambda \sim r_1$ (where $r_1$ is the position where the buoyancy force vanishes $dp/dr = 0$) if [30]

$$\frac{r - r_1}{r_1} \geq \frac{\rho_p}{\rho}. \tag{18}$$

For the example $\rho_p / \rho \cong 0.1$ one would have $r > 1.1 r_1$, which shows that a small penetration into the vortex core wall is sufficient to assure stability.

Figure 4 shows the results of the linear analysis on the kink instability for the constant current density equilibrium profile with azimuthal velocity $0.1 v_A / \sqrt{0.01 + r^2}$ and axial velocity $v_{0z} v_A r^2 / R^2$. The constant current density equilibrium profile has the following form:

$$B_\theta = B_R r / R; \quad p = p_0 \left(1 - \frac{r^2}{R^2}\right). \tag{19}$$

Here $p$ is the pressure, $B_\theta$ is the azimuthal magnetic field, $B_R$ is the value of $B_\theta$ at the edge of the plasma cylinder ($r = R$) and $R$ is the radius of the pinch. The kink instability growth rates are reduced to zero for $v_{0z} > 1$ [24]. We also present results for the constant electron velocity (Bennett) equilibrium profile. The Bennett equilibrium profile (with $a = 2 r_0 / 3$ for this paper) has the following form:

$$B_{0\theta} = \sqrt{2}\left(a+\frac{1}{a}\right)\frac{r/a}{1+r^2/a^2}; \quad P_0 = \left(a+\frac{1}{a}\right)^2 \frac{1}{\left(1+r^2/a^2\right)^2}; \quad P_0 = n_0 \qquad (20)$$

Figure 5 shows the results for the Bennett equilibrium profile with the same velocity as in figure 4. The instability growth rates are reduced to zero for $v_{0z} > 1.2$.

It has been proposed to produce sheared helical flow with a corrugated z-pinch load [30]. For this we consider a capillary tube which possesses a helical saw tooth corrugated surface (see figure 6). A pinch discharge along the surface of the tube produces jets emitted from the inner corner of the wedges, with the jets supersonically propagating along the tube surface stabilizing it against the *m* = 0 and *m* = 1 magnetohydrodynamic instabilities. In addition, the helical winding generates a rotational flow suppressing the Rayleigh-Taylor instability. To reach the highest possible densities, the compression must be isentropic, requiring a current pulse with a properly chosen time dependence [3]. After the capillary tube is compressed, with the DT reaching its highest density, the DT is ignited with a laser pulse at one point launching a thermonuclear detonation wave from this point.

If the electrical conductivity of the capillary tube is sufficiently high, and the implosion short compared to the time needed for the magnetic field to diffuse into the tube, the magnetic field cannot penetrate into the tube. With the electric current flowing along the surface of the capillary tube, the magnetic body force $(1/c)\mathbf{j}\times\mathbf{B}$ is directed perpendicularly onto the surface of the saw tooth (see figure 6). The magnetic field then acts like a piston onto the surface of the capillary tube, and one has to set the wedge implosion velocity $v_0$ equal to the Alfvén velocity $v_A = B/\sqrt{4\pi\rho}$, where $B = 0.2 I/r_0$ is the magnetic field of the current *I* in Ampere, with $r_0$ the initial radius of the capillary tube and $\rho$ its density. As a result, sheet like jets are ejected along the imploding saw tooth. The recoil from these jets generates a massive "slug" moving in a direction opposite to the jet. Both the jet and slug generate shear and implode the capillary tube.

## V.  CONCLUSION

There are two advantages of the fast-ignition method with isentropic compression over the method of heating the entire plasma column.  In the fast-ignition method most of the plasma can remain much cooler.  There only the hot-spot needs to be raised to the ignition temperature, with a programmed current pulse isentropically compressing the plasma on a lower adiabat.  The sound speed is higher in a hot plasma and with it the instability growth rate.  This means that a lower sheared flow velocity can reach the Mach number needed to stabilize the plasma.  Thus stabilizing the plasma with the fast-ignition method with isentropic compression takes less energy and is more efficient than heating the entire z-pinch plasma to thermonuclear temperatures.

## VI.  ACKNOWLEDGEMENT

This work has been supported in part by the U. S. Department of Energy under Grant No. DE-FG02-06ER54900.

**Figure Captions**

Figure 1  Pinch instability due to $\nabla \times \mathbf{B} = \mathbf{0}$, and vortex stability due to $\nabla \times \mathbf{v} = \mathbf{0}$ and the Bernoulli theorem.

Figure 2  This graph compares the kink instability growth rates for different functional forms of azimuthal flow.  When $v_{0\theta} = 0$ there is no azimuthal flow.  Sheared azimuthal flow in the approximate form of a vortex ($v_\theta = 0.1/\sqrt{0.01 + r^2}$) reduces $m = 1$ growth rates more efficiently than the other two azimuthal flow profiles.

Figure 3  Opposing solenoids in series with a wire array.

Figure 4  This is a 3 dimensional plot of the kink instability growth rate as a function of the dimensionless axial wave number ($k_z r_0$) and the Mach like parameter $v_{0z}$.  As an example of how to interpret this graph consider the kink instability growth rate for $v_{0z} = 0.6$.  The growth rate is zero for $k_z r_0 = 2$.  It increases to a maximum of ~ 0.15 ($v_A$ / $r_0$) for $k_z r_0 \approx 8$ then drops to zero for $k_z r_0 > 10$.  The kink instability growth rates are negligible for all wave lengths analyzed here ($1 < k_z r_0 < 20$) for $v_{0z} > 1.0$.

Figure 5 shows the results for the Bennett equilibrium profile with the same velocity as in figure 4.  The instability growth rates are reduced to zero for $v_{0z} > 1.2$.

Figure 6  This drawing illustrates a Z-pinch load with helical saw tooth threading. Current flowing along the surface produces a magnetic force $\propto \mathbf{j} \times \mathbf{B}$ that collapses the threads. A helical sheet-like jet is produced that generates velocity shear as it flows along the imploding wire.

Figure 1

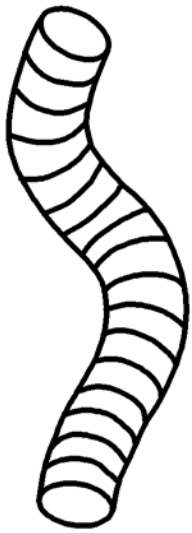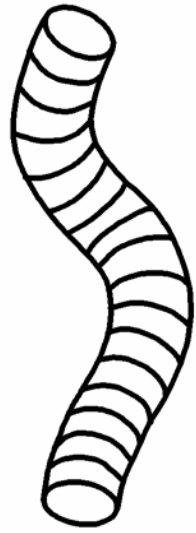

$\nabla \times \mathbf{B} = 0$
unstable

$\nabla \times \mathbf{v} = 0$
stable

Figure 2

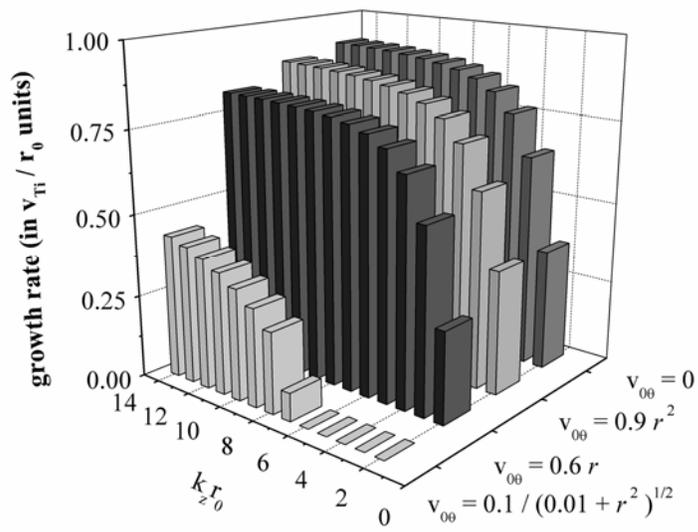

Figure 3

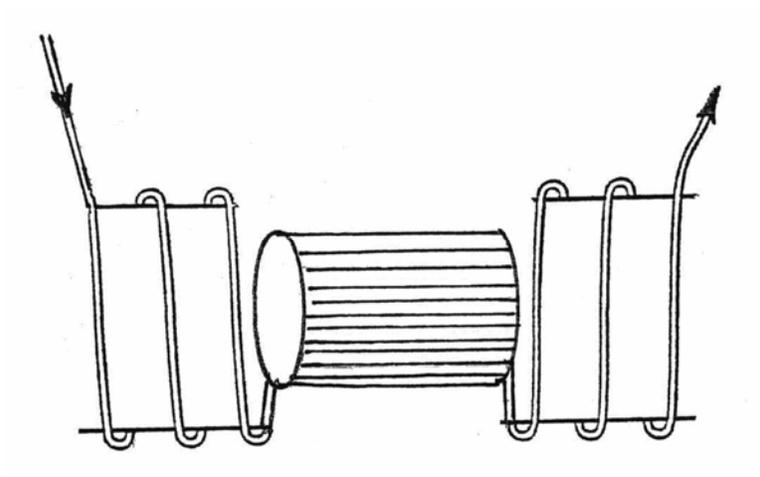

Figure 4

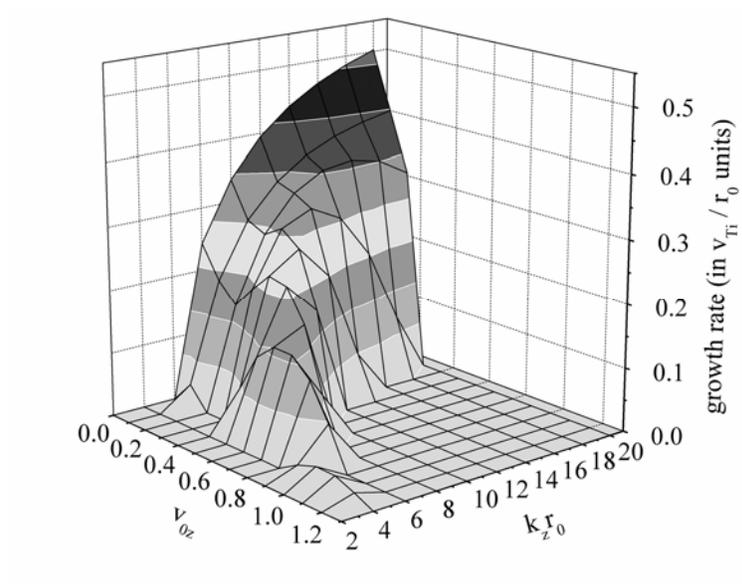

Figure 5

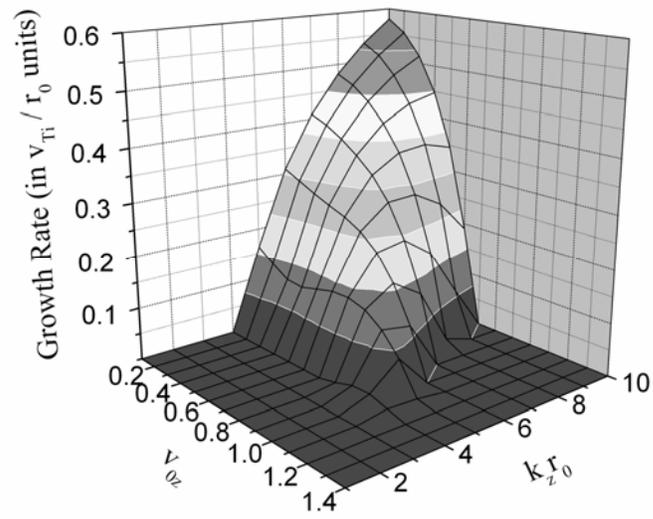

Figure 6

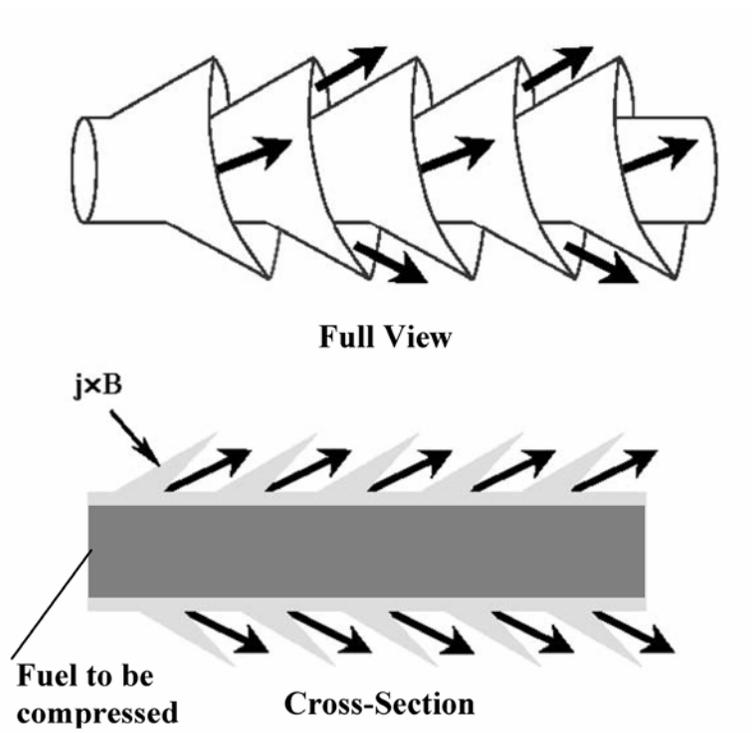

**Full View**

**Cross-Section**

j×B

Fuel to be compressed